\documentclass[prb,twocolumn,groupeaddress]{revtex4-2}

\usepackage{amsmath}
\usepackage{graphicx}
\usepackage[colorlinks=true, allcolors=blue]{hyperref}
\usepackage{braket}
\usepackage{physics}

\begin{document}

\title{Correlation-assisted spin-selective metallicity in strained and bilayer altermagnets}
\author{Marnin J. Di Nunzio}
\affiliation{Theoretische Physik III, Ruhr-Universit\"at Bochum,
  D-44780 Bochum, Germany}
\author{Frank Lechermann}
  \affiliation{Theoretische Physik III, Ruhr-Universit\"at Bochum,
  D-44780 Bochum, Germany}
\author{Ilya M. Eremin}
\affiliation{Theoretische Physik III, Ruhr-Universit\"at Bochum,
  D-44780 Bochum, Germany}

\begin{abstract}
Altermagnets combine vanishing net magnetization with a momentum-dependent spin-split electronic structure, providing a route to spin-polarized carriers without ferromagnetism. Here, we investigate how electronic correlations, doping, uniaxial strain, and interlayer coupling control altermagnetism within a minimal Hubbard-type model for mono- and bilayer systems. Comparing Hartree--Fock theory with the rotationally invariant slave-boson (RISB) approach, we demonstrate the robustness of altermagnetic order against quasiparticle renormalization and reveal a separation between the onset of magnetic order and the loss of quasiparticle coherence at stronger coupling. In the monolayer, doping produces a pronounced particle--hole asymmetry, while its combination with uniaxial strain generates a fully spin-polarized Fermi surface close to half-filling. In the bilayer, stacking that favors ferroic alignment of the layer altermagnetic order parameters allows their momentum-dependent spin splittings to combine constructively. Weak asymmetric doping additionally induces intra-unit-cell charge order, resulting in fully spin-polarized low-energy carriers. At a bilayer filling of five electrons in four orbitals, we find a continuous paramagnet-to-altermagnet transition followed, at stronger coupling, by an evolution toward a Mott-like regime with strongly suppressed quasiparticle weight. Our results establish correlations, strain, doping, and stacking as complementary means of controlling altermagnetic metals and generating fully spin-polarized Fermi surfaces.
\end{abstract}

\maketitle

\section{Introduction}

Altermagnetism has recently emerged as a distinct form of collinear magnetic order that combines important characteristic properties of conventional ferromagnets and antiferromagnets \cite{Smejkal2022PRX, Smejkal2022PRXSpin}. In an altermagnet, oppositely polarized magnetic sublattices compensate each other, resulting in a vanishing net magnetization, while being related by crystal rotations rather than by translations or inversion. As a consequence, the electronic bands can exhibit a sizable non-relativistic, momentum-dependent spin splitting despite the absence of a macroscopic magnetic moment \cite{Smejkal2022PRX}. The corresponding spin polarization typically displays an even-parity $d$-, $g$-, or higher-wave symmetry in momentum space. This combination makes altermagnets attractive both as a platform for unconventional magnetic phenomena and for spintronic applications, where spin-polarized currents without the stray fields associated with ferromagnets are particularly desirable. Recent experimental progress, including direct observations and spatial imaging of altermagnetic signatures, has further stimulated the search for mechanisms to generate and control altermagnetic electronic states \cite{Krempasky2024,Zhu2024,Amin2024}.

An important question is therefore how the altermagnetic state and, in particular, its spin-split Fermi surface can be manipulated by experimentally accessible control parameters. Strain provides a natural route because the altermagnetic spin splitting is intrinsically tied to crystal symmetry and anisotropy. Recent studies have demonstrated that lattice deformations can modify the magnitude and symmetry of altermagnetic spin splitting and can induce transitions between distinct magnetic states \cite{Li2025,Khodas2026,Guo2026,Forte2026,Guo2026}. Another promising control parameter is dimensionality. In two-dimensional systems, suitable stacking of otherwise conventional magnetic layers can generate the symmetry relations required for altermagnetism \cite{Pan2024,Zeng2024,Sun2024}. Interlayer hopping and slight electron/hole doping then provides an additional microscopic energy scale for reshaping the spin-resolved bands, suggesting that bilayer systems may offer substantially more flexibility than their monolayer counterparts.

While the symmetry principles and mean-field electronic structure of altermagnets are by now well established, the role of electronic correlations beyond static mean-field theory has received comparatively less attention~\cite{ferrari24,sato24,leeb24,bose25,giuli25}. Recent studies employing dynamical mean-field theory and density-matrix renormalization group have demonstrated that correlations can substantially affect the spectral properties and competing phases of microscopic altermagnetic Hubbard models\cite{DelRe2025,He2025}. This raises the question to what extent the mean-field altermagnetic phase and its characteristic spin-split Fermi surfaces remain robust once quasiparticle renormalization is taken into account.

This issue is particularly relevant close to a correlation-driven metal--insulator transition, where quasiparticle (QP) renormalization can strongly modify the low-energy band structure and therefore the altermagnetic spin splitting. Hubbard-type models provide a minimal setting in which altermagnetism can be generated and its interplay with interaction, doping, and kinetic anisotropy can be studied systematically \cite{Das2024,Dong2026,Li2026}. Going beyond conventional Hartree--Fock theory is then essential for determining to what extent the resulting phase diagram and Fermi-surface topology survive strong quasiparticle renormalization.

Of particular interest in this context is the possibility of realizing fully spin-polarized Fermi surface with a small net magnetization. Such a state combines near magnetic compensation with $100\%$ spin polarization of the low-energy charge carriers and is therefore conceptually distinct from conventional ferromagnetic half-metals. Recent material-specific calculations have demonstrated that antiferromagnetic half-metallicity can occur in sliding bilayer altermagnets \cite{Zhang2026}. This raises the more general question of whether compensated half-metallicity can emerge naturally from the interplay between electronic correlations, lattice anisotropy, doping, and interlayer hybridization in a minimal microscopic model.

In this work, we address this question using a minimal Hubbard-type model for mono- and bilayer altermagnets. We first establish the magnetic phase diagram within static mean-field theory and subsequently include correlation-induced quasiparticle renormalization using the rotationally invariant slave-boson (RISB) mean-field formalism \cite{li89,lec07}. For the monolayer, we show that the combined action of uniaxial strain and doping drives the altermagnetic metal into a compensated half-metallic regime close to the metal--insulator transition, where only one spin species contributes to the Fermi surface with small net magnetization. We then consider two distinct bilayer stackings and demonstrate that interlayer hybridization provides an alternative route to the same state. Remarkably, in the bilayer geometry doping alone can be sufficient to generate a fully spin-polarized metallic Fermi surface due to additional inraunit-cell charge order, which also lifts the total compensation of the magnetic moment in altermagnet. Our results thereby identify strain, quasiparticle renormalization, and interlayer coupling as complementary microscopic ingredients for controlling compensated half-metallicity in correlated altermagnets.




\section{Theoretical Approach}

\begin{figure*}[htp]
      \includegraphics[width=\textwidth]{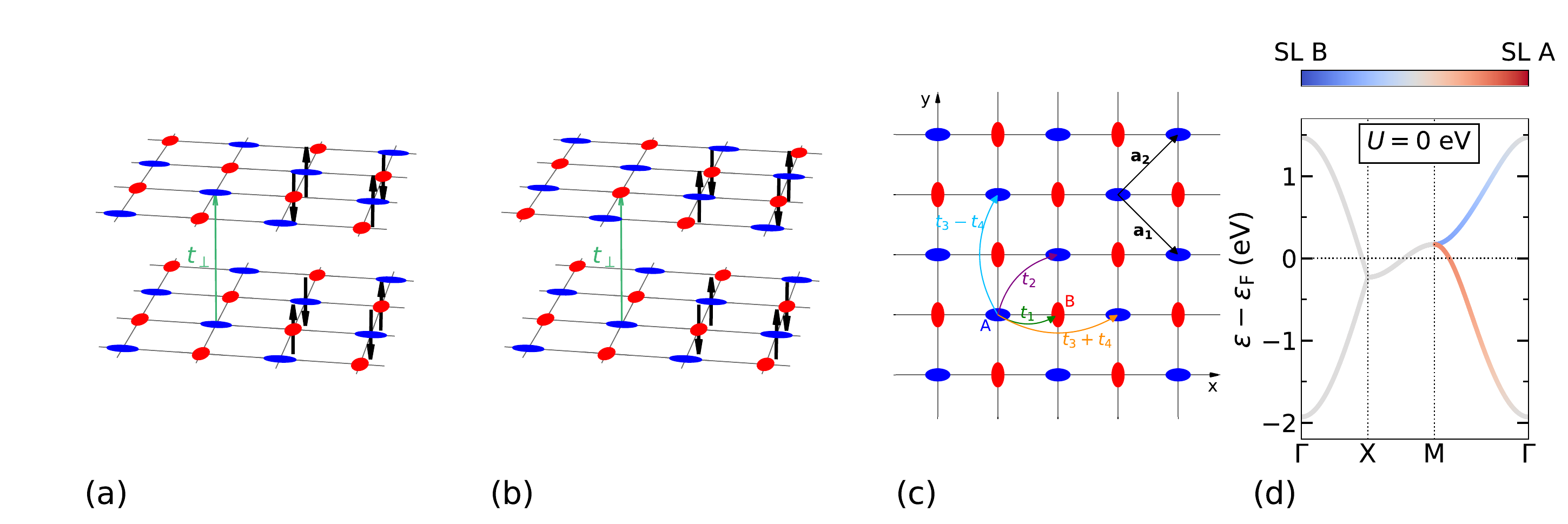}
      \caption{Illustration of the altermagnetic bilayer model with (a) AA and (b) AB stacking of the two layers. The arrows indicate the antiferromagnetic arrangements of the spins within a antiferromagnetic plaquette. (c) Structure of an individual layer, consisting of a square bipartite lattice with A (blue) and B (red) sublattices. The hopping amplitudes $t_1$, $t_2$, $t_3$, and $t_4$ describe the relevant intra- and inter-sublattice hopping processes. (d) Tight-binding dispersion of the single-layer case for chosen hoppings~\cite{Gondolf2025} (see text), colored according to the sublattice orbital weight.
        } \label{fig1}
\end{figure*}	

The roadmap of our model-Hamiltonian is to start from a single-layer model and advance thereon by building up bilayer models, as sketched in Fig.~\ref{fig1}. Let us first focus on the single-layer case with a checkerboard arrangement of $A$ and $B$ sites and one orbital per site, respectively.
Using a basis of atomic-like orbitals, ($\Psi_{\mathbf{k}}=(c_{A\mathbf{k}},c_{B\mathbf{k}})^T$), the kinetic part of the single-layer Hamiltonian can be written in reciprocal space as~\cite{Roig2024,Gondolf2025}
\begin{equation}
{\cal H}_{\rm kin} ({\bf k})=\left(\begin{array}{rrr}
H_{AA}({\bf k}) & H_{AB}({\bf k})\\
H_{BA}({\bf k}) & H_{BB}({\bf k})\\
\end{array}\right)\qquad,
\end{equation}
The individual matrix elements are given by
\begin{eqnarray}
H_{AA}({\bf k})&=&-2\,t_2\,(\cos k_x + \cos k_y)-4\,t_3\,\cos k_x\,\cos k_y \nonumber\\
&&-4\,t_4\,\sin k_x\,\sin k_y\\
H_{BB}({\bf k})&=&-2\,t_2\,(\cos k_x + \cos k_y)-4\,t_3\,\cos k_x\,\cos k_y \nonumber\\
&&+4\,t_4\,\sin k_x\,\sin k_y\\
H_{AB}({\bf k})&=&
-2\,t_1\,\left(\cos\frac{k_x+k_y}{2}+\cos\frac{k_x-k_y}{2}\right)\\
&=&H_{BA}({\bf k})\nonumber
\end{eqnarray}
Here, $t_1$ describes hopping between the $A$ and $B$ sublattices, whereas $t_2$, $t_3$, and $t_4$ correspond to hopping processes within the same sublattice, as indicated schematically in Fig.~\ref{fig1}(c). The $t_2$ and $t_3$ contributions are identical for the two sublattices and therefore enter the diagonal elements $H_{AA}$ and $H_{BB}$ with the same sign. In contrast, the $t_4$ term changes sign between the $A$ and $B$ sublattices. This sublattice-dependent contribution is essential for generating the momentum-dependent distinction between the two sublattices that underlies the altermagnetic band structure
\begin{equation}
\varepsilon_0(\mathbf{k})\tau_0
+\varepsilon_x(\mathbf{k})\tau_x
+\varepsilon_z(\mathbf{k})\tau_z\quad,
\end{equation}
where $\varepsilon_0(\mathbf{k})=
-2t_2(\cos k_x+\cos k_y)
-4t_3\cos k_x\cos k_y$, $
\varepsilon_x(\mathbf{k})=
-2t_1\left[
\cos\frac{k_x+k_y}{2}
+\cos\frac{k_x-k_y}{2}
\right]$, and $\varepsilon_z(\mathbf{k})=
-4t_4\sin k_x\sin k_y$.
This representation makes the different roles of the hopping processes particularly transparent. While $\varepsilon_0(\mathbf{k})$ determines the sublattice-independent part of the dispersion and $\varepsilon_x(\mathbf{k})$ hybridizes the $A$ and $B$ states, $\varepsilon_z(\mathbf{k})$ distinguishes the two sublattices in a momentum-dependent manner. In particular, $\varepsilon_z(\mathbf{k})$ changes sign under a (90$^\circ$) rotation, $k_x\rightarrow-k_y$, $k_y\rightarrow k_x$, and therefore possesses $d_{xy}$-type momentum-space symmetry. This property becomes important once opposite exchange fields are introduced on the two sublattices, resulting in the characteristic momentum-dependent spin splitting of an altermagnet $\varepsilon_0(\mathbf{k})
\pm
\sqrt{\varepsilon_x^2(\mathbf{k})+\varepsilon_z^2(\mathbf{k})}$. For the actual values of the hopping amplitudes we utilized the
parametrization by Gondolf {\sl et al.}~\cite{Gondolf2025}, i.e. $\{t_1,t_2,t_3,t_4\}=\{0.425, 0.05, -0.025, -0.075\}$\,eV. 
The resulting basic single-layer dispersion for the half-filled case of one electron per site is depicted in Fig.~\ref{fig1}(d). Close to the $\Gamma$-point, it shows the dominant bonding-antibonding splitting originating from the two orbitals in the unit cell. Note the 
decisive sublattice-selective character of the dispersions along the Brillouin-zone diagonal $\Gamma$$-$M, which connects to
the real-space checkerboard pattern. It will be seen that this part of the dispersion becomes susceptible to altermagnetic
splitting.

The bilayer models are constructed from two identical copies of this single-layer Hamiltonian supplemented by an interlayer hopping $t_\perp$. The precise form of the latter depends on whether the two layers are arranged in the $AA$ or $AB$ stacking configuration [Figs.~\ref{fig1}(a,b)]. As demonstrated by the dispersions in Fig.~\ref{fig3}, the stacking geometry therefore provides an additional degree of freedom that modifies the hybridization between the layers and, consequently, the resulting quasiparticle band structure.

To investigate the effect of electronic correlations on the altermagnetic state, we supplement the tight-binding Hamiltonian by an on-site intraorbital Hubbard interaction,
\begin{equation}
\mathcal{H}_{\mathrm{int}}
=
U\sum_{i,\mu}
n_{i\mu\uparrow}n_{i\mu\downarrow},
\label{eq}
\end{equation}
where $i$ labels the unit cell and $\mu$ denotes the sublattice degree of freedom. In the following we analyze the same interacting model within two complementary mean-field approaches. We first employ a conventional Hartree--Fock (HF) decoupling, which was previously used for describing the altermagnetic order~\cite{Roig2024} and provides a transparent description of the onset and spatial structure of the magnetic order. We subsequently compare these results with calculations within the rotationally invariant slave-boson (RISB) framework, which additionally incorporates local correlation-induced quasiparticle renormalizations. This comparison allows us to determine to what extent the altermagnetic solutions and their stability obtained within static Hartree--Fock theory survive beyond the un-renormalized single-particle picture especially near the metal-insulator transition.

Within the Hartree--Fock treatment, the local Hubbard interaction is decoupled in the density and magnetic channels, following the approach of Ref.~\cite{Gondolf2025}. For a given sublattice index $\mu=A,B$ one obtains 
\begin{align}
    \mathcal{H}_\mathrm{int}^{(\mu)} & = U n_{\mu \uparrow} n_{\mu\downarrow}\notag\\
    &\approx \sum_{\sigma} \frac{U}{2} \bigl( \expval{n_\mu} - \sigma N_\mu \bigr) n_{\mu\sigma} - U \expval{n_{\mu\downarrow}} \expval{n_{\mu\uparrow}},
\end{align}
where $\expval{n_\mu} = \expval{n_{\mu\uparrow} + n_{\mu\downarrow}}$ denotes the local charge density, while $N_\mu = \expval{n_{\mu\uparrow} - n_{\mu\downarrow}}$ is the local magnetic order parameter. In the altermagnetic state, the latter acquires opposite signs on the two sublattices, while the total magnetization remains compensated. The resulting mean-field Hamiltonian
\begin{equation}
    \mathcal{H}_{\mathrm{MF, int}}^{(\mu)}= \sum_{\sigma} \frac{U}{2} \bigl(\expval{\delta n_\mu} - \sigma N_\mu \bigr) n_{\mu \sigma}.
\end{equation}
is solved by numerical diagonalization. The eigenvalues and eigenvectors are subsequently used to recalculate the local charge densities and magnetic order parameters as thermal expectation values at T=0.01$\mathrm{eV}$. The chemical potential is adjusted in each iteration to keep the total filling fixed. This procedure is iterated until self-consistency is reached. The resulting HF solution provides our reference description of the altermagnetic state and, in particular, of its stability as a function of interaction strength and the remaining model parameters.

The analytic expression of the mean field energies, for which a constant term containing the filling was absorbed into the chemical potential under the assumption of $\expval{n_{\mathrm{A}}} = \expval{n_{\mathrm{B}}}$, looks as follows
\begin{align}
    E_{\sigma}^{\beta}(\mathbf{k})&=\varepsilon_{0}(\mathbf{k})-\frac{U}{4}\sigma(N_A+N_B)\notag\\
    &+ \beta \sqrt{|\varepsilon_{x}(\mathbf{k})|^2+(\varepsilon_{z}(\mathbf{k})-\frac{U}{4}\sigma(N_A-N_B))^2}.
\end{align}
$\beta$ denotes the upper ($\beta=+1$) and lower ($\beta=-1$) bands. 

To examine whether the Hartree--Fock results remain robust when local many-body renormalizations are taken into account, we additionally solve the interacting model within the RISB framework at the saddle-point level~\cite{li89,lec07,bunemann07,isi09,pie18,fac18}. Again importantly, the purpose of the RISB calculation here is not to introduce a different microscopic interaction model: the same Hubbard interaction is considered in both approaches. The comparison therefore directly probes the sensitivity of the predicted altermagnetic state to the approximation used for treating electronic correlations.

In RISB, the physical electron operator is represented in terms of auxiliary quasiparticle (QP) fermions and slave-boson amplitudes.  In essence, this many-body technique builds up on a fragmentation of the QP character and the local-excitation character on the operator level, here for orbital character $m$ formally reading
\begin{equation}
c_{m\sigma}^\dagger=R(\{\phi\})\,f_{m\sigma}^\dagger\qquad.
\end{equation}
The $f$ degree of freedom describes the fermionic QP state, and the bosonic set $\{\phi\}$ provides access to the local multiplets. The $R$ function describes the QP renormalization from correlations. In the generic RISB approach, there is one
$\phi$ boson for each pair of local Slater determinants
$(|\{n^{(p)}_{m\sigma}\}\rangle,|\{n^{(p')}_{m'\sigma'}\}\rangle)=:
(|{\cal A}_p\rangle,|{\cal Q}_{p'}\rangle)$, with $p,p'$ denoting particle sectors. But as we
here focus on normal-state properties, there are only bosons connecting states in
identical particle sectors $p=p'$ considered. The extension to the explicit
superconducting phase is discussed in Ref.~\onlinecite{isi09}. Note that formally, the
index ${\cal A}$ labels a local atomic state and ${\cal Q}$ the QP degree of
freedom~\cite{lec07}.

The RISB electronic self-energy $\Sigma$ is local and consists of a term linear in frequency as well as a static part, i.e.
\begin{equation}
\mathbf{\Sigma}(\omega)=\omega\left(1-\mathbf{Z}^{-1}\right)
+{\bf \Sigma}^{\rm stat}\,\,,\label{eq:Sigma_physical1}
\end{equation}
whereby $\mathbf{Z}=\mathbf{R}\mathbf{R}^\dagger$ is the QP-weight matrix in orbital-spin space. While the former part gives rise to Fermi-liquid properties of the correlated state, the latter describes the renormalization of local
hybridizations and crystal fields. There are no frequency-dependent Hubbard bands in the mean-field RISB scheme, however, useful information about the local excitations (e.g.
average multiplet occupations, spin correlations, etc.) within a correlated metal is still included by the frequency-independent bosonic amplitudes. The method hence
lacks the full-frequency dependence of the dynamical mean-field theory (DMFT) self-energy, but is still well suited (here at formal $T=0$) for a wide class of correlated materials
problems. For instance, it naturally may describe local-moment formation from $\langle S^2\rangle>0$, whereas e.g. static correlation schemes can only access ordered
moments with $\langle S\rangle\neq 0$. For further methodological details see e.g. Refs.~\onlinecite{lec07,pie18}.

Note that in the bilayer studies, we discuss here cluster-RISB~\cite{lec07,ferrero09} results, with the two-site cluster connecting sites across the
bilayer along the basic hopping path $t_\perp$ as the basic cellular cluster. Thus, interlayer self-energies are included in these correlation studies, while intralayer correlations are still treated on the single-site level.

\section{Results}

\subsection{Square-lattice with strain}

\begin{figure}[t]
      \includegraphics[width=8.5cm]{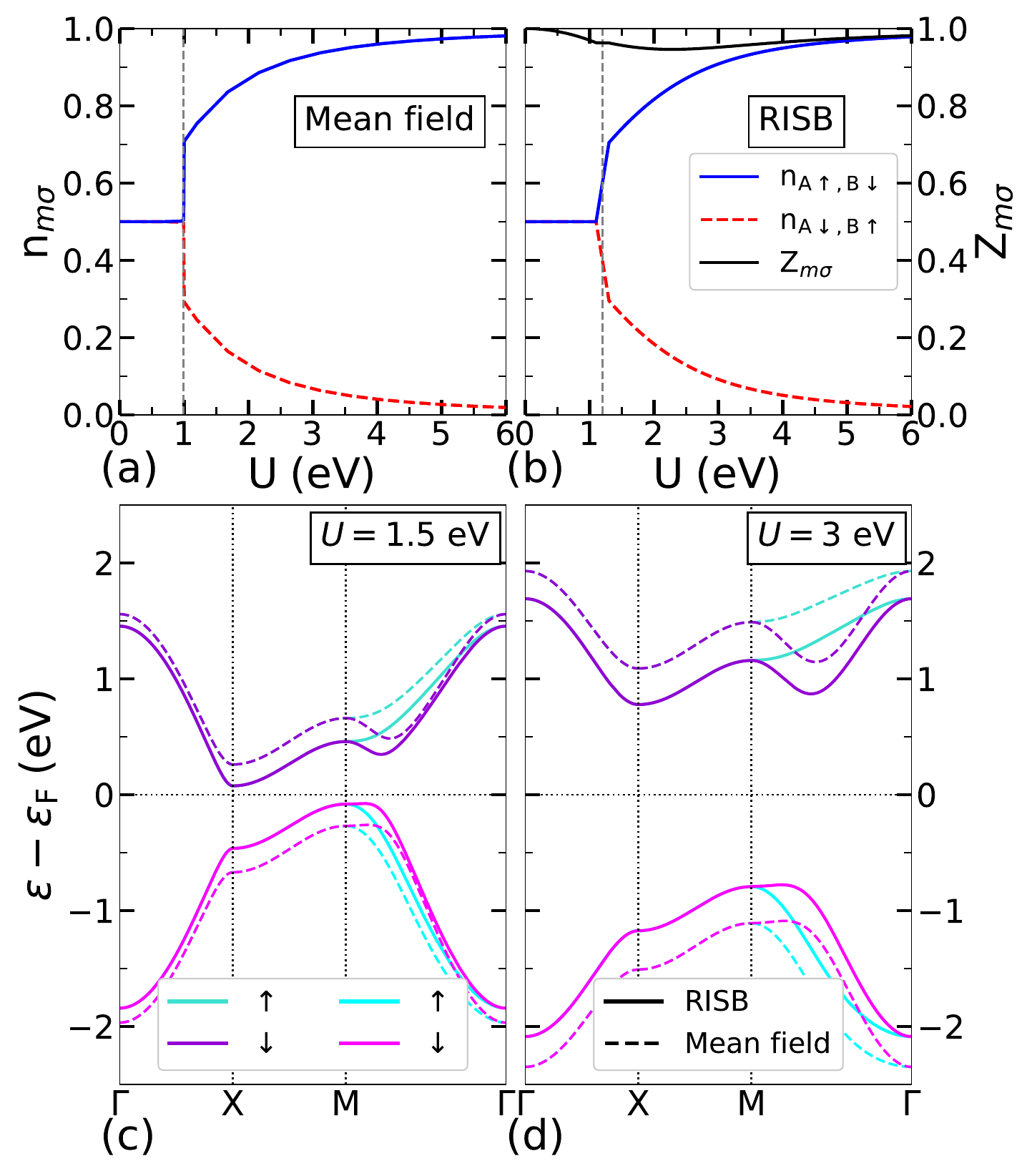}
      \caption{Comparison of the Hartree--Fock mean-field and rotationally invariant slave-boson (RISB) descriptions of the altermagnetic transition at half-filling $n=2$ (2 atoms per unit cell). (a) Spin- and sublattice-resolved occupations \(n_{A\uparrow,B\downarrow}\) and \(n_{A\downarrow,B\uparrow}\) as a function of the Hubbard interaction \(U\) within Hartree--Fock theory. The vertical dashed line indicates the critical interaction \(U_c\) for the onset of altermagnetic order. (b) Corresponding RISB results together with the QP weight \(Z_m\) (black line). The finite \(Z_m\) across the magnetic transition demonstrates that altermagnetic order develops within the correlated metallic regime. (c,d) Quasiparticle dispersions along the high-symmetry path \(\Gamma-X-M-\Gamma\) for \(U=1.5\,\mathrm{eV}\) and \(U=3\,\mathrm{eV}\), respectively, comparing RISB (solid curves) and Hartree--Fock (dashed curves). Colors distinguish the two spin sectors \(\sigma=\pm1\). The comparison illustrates the correlation-induced QP renormalization while preserving the characteristic momentum-dependent altermagnetic spin splitting.} \label{fig2}
\end{figure}

We first assess the robustness of the mean-field altermagnetic state against correlation-induced QP renormalization and compare the Hartree--Fock results with the RISB approximation in Fig.~\ref{fig2}. At half-filling ($n=2$ as we have 2 atoms per unit cell), both approaches yield a transition from the paramagnetic state, characterized by equal occupations of the two spin-sublattice sectors, to the compensated altermagnetic state with \(n_{A\uparrow,B\downarrow}\neq n_{A\downarrow,B\uparrow}\). The transition occurs at a finite critical interaction \(U_c\), whose value is very similar in both approaches. Importantly, the RISB quasiparticle weight \(Z_m\) remains finite, spin-independent and close to 1 throughout the ordered regime, demonstrating that the altermagnetic transition remains Slater-like also within RISB without possible localization of the QPs, i.e. there is no Mott physics developing in this case. At the same time the correlations beyond mean-field reduce the size of the magnetic gap as seen in the QP dispersion, which is illustrated in Figs.~\ref{fig2}(c) and (d) for \(U=1.5\) and \(3\,\mathrm{eV}\), respectively. Both approaches nevertheless retain the same characteristic momentum-dependent altermagnetic spin splitting and overall band topology. Thus, while QP correlations quantitatively renormalize slightly the critical interaction and electronic dispersion, the comparison establishes that the altermagnetic state obtained at the Hartree--Fock level remains robust within RISB and the metal-insulator transition is Slater-like. 
%

\begin{figure}[t]
      \includegraphics[width=8.5cm]{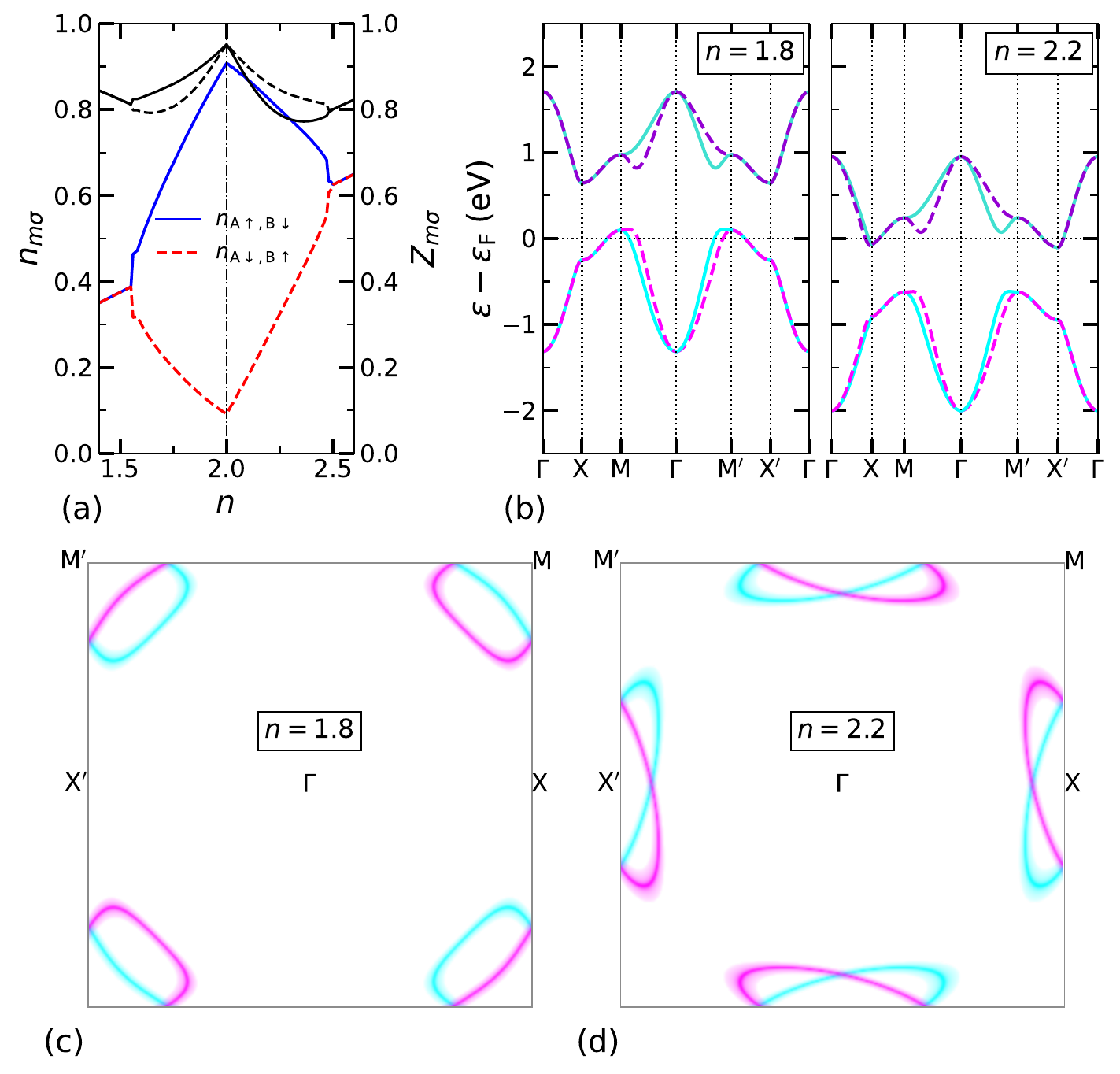}
      \caption{Calculated doping dependence of the correlated altermagnetic state within RISB. (a) Spin- and sublattice-resolved occupations \(n_{A\uparrow,B\downarrow}\) and \(n_{A\downarrow,B\uparrow}\), together with the spin-resolved quasiparticle weights \(Z_{m\sigma}\), as a function of the electronic filling \(n\). Half-filling corresponds to \(n=2\). The QP weight displays a pronounced asymmetry between different spin species away from half filling. (b) Renormalized QP dispersions along the high-symmetry path for representative hole- (\(n=1.8\)) and electron-doped (\(n=2.2\)) cases. Cyan and magenta denote the two spin sectors. (c,d) Corresponding spin-resolved Fermi surfaces for \(n=1.8\) and \(n=2.2\), respectively, illustrating the distinct Fermi-surface reconstruction on the two sides of half-filling.} \label{fig3}
\end{figure}

Having established the stability of the altermagnetic solution against QP renormalization at half-filling, we next investigate its evolution away from half-filling. Figure~\ref{fig3} summarizes the RISB results as a function of the electronic filling \(n\). The altermagnetic state remains stable over a sizable doping range on both sides of half-filling.  Upon doping, the Slater-like insulating state evolves into a correlated metal. At the same time, the QP weight \(Z_{m\sigma}\) decreases and develops a pronounced spin asymmetry away from half-filling. At the same time, it exhibits a pronounced spin asymmetry away from the half-filling. In particular, the renormalization of electron and hole doping lead to distinctly different evolutions of the correlation-induced band renormalization, despite a qualitatively similar suppression of the altermagnetic order away from \(n=2\). This asymmetry is directly reflected in the QP dispersions shown in Fig.~\ref{fig3}(b) for representative hole- and electron-doped fillings \(n=1.8\) and \(n=2.2\), respectively. Although the characteristic momentum-dependent spin splitting of the altermagnetic state persists in both cases, the position of the Fermi level relative to the spin-split bands evolves differently on the two sides of half-filling. Consequently, the corresponding Fermi surfaces [Figs.~\ref{fig3}(c,d)] undergo markedly different reconstructions: hole doping produces spin-split pockets centered near the Brillouin-zone corners, whereas electron doping yields elongated pockets along the zone boundaries. The combined evolution of the QP weight and Fermi-surface topology therefore demonstrates that doping provides an efficient means of controlling both the strength of correlations and the spin-polarized low-energy electronic structure of the altermagnetic metal. This spin-dependent difference in the QP renormalization should be visible in ARPES. This spin-dependent QP renormalization should, in principle, be observable in spin-resolved ARPES, with the relative quasiparticle residues of the \(A\uparrow/B\downarrow\) and \(A\downarrow/B\uparrow\) sectors reversing upon changing the sign of doping, which would be interesting to test experimentally. 
%

This doping-induced redistribution of the spin-split QP bands makes the system particularly susceptible to symmetry-lowering perturbations, allowing uniaxial strain to eliminate one spin sector from the Fermi surface already at very small deviations from half-filling. We model the uniaxial strain by modifying the terms, which contain the hopping parameters $t_1$,$t_3$ and $t_4$, as follows
\begin{equation}
\varepsilon_0(\mathbf{k},\epsilon)=\varepsilon_0(\mathbf{k})+4t_3 \epsilon\sin k_x\sin k_y,
\end{equation}
\begin{equation}
    \varepsilon_x(\mathbf{k},\epsilon) = \varepsilon_x(\mathbf{k})+4t_1 \epsilon\sin\frac{k_x}{2}\sin\frac{k_y}{2},
\end{equation}
\begin{equation}
    \varepsilon_z(\mathbf{k},\epsilon)=\varepsilon_z(\mathbf{k})-4t_4\epsilon\cos k_x\cos k_y.
\end{equation}
Here $\epsilon$ represents the strain parameter.
%
\begin{figure}[b]
      \includegraphics[width=8.5cm]{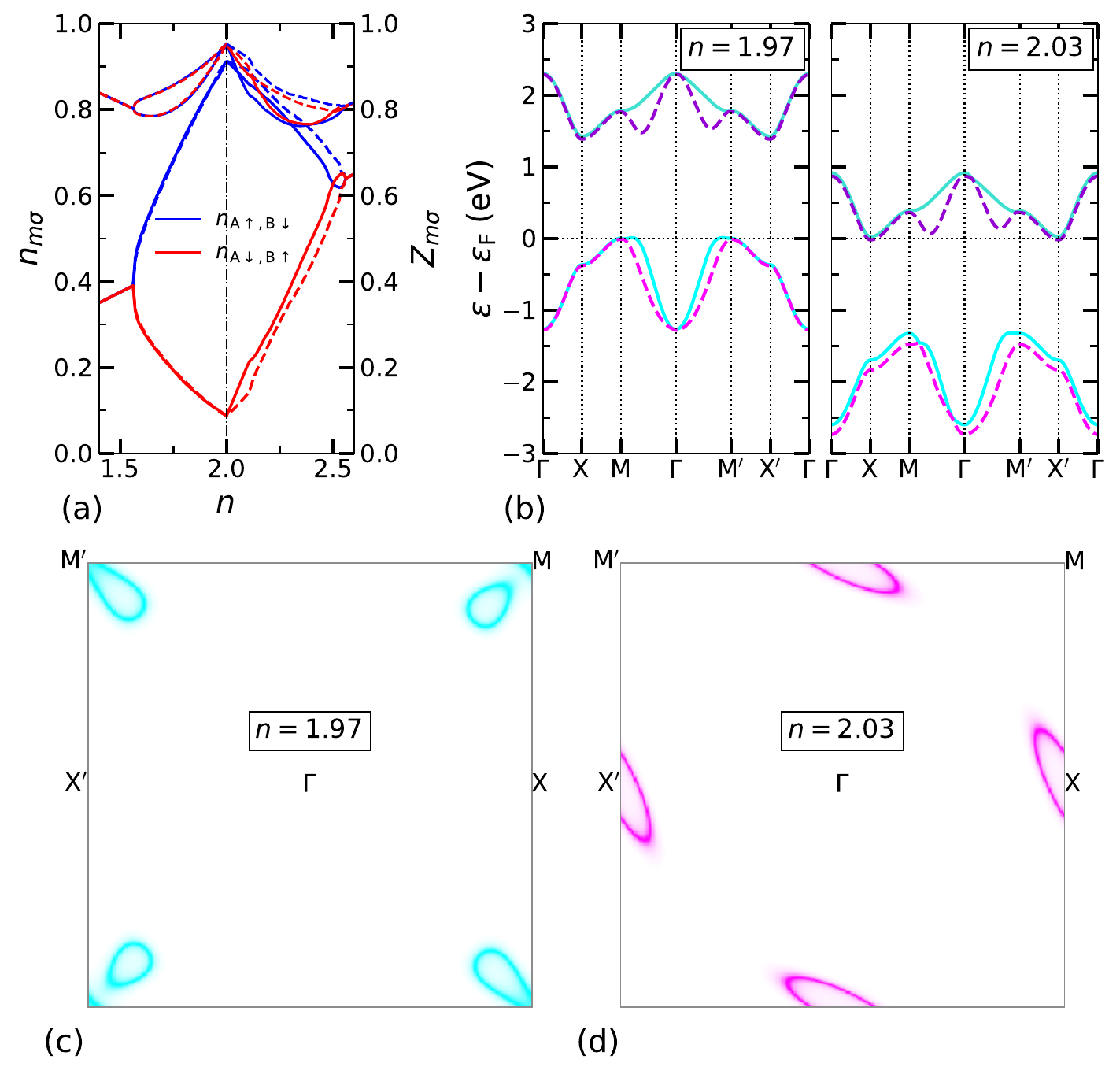}
      \caption{Calculated strain-induced fully spin-polarized metallic state close to half-filling. (a) Spin- and sublattice-resolved occupations \(n_{A\uparrow,B\downarrow}\) and \(n_{A\downarrow,B\uparrow}\), together with the spin-resolved QP weights \(Z_{m\sigma}\), as a function of filling \(n\) in the presence of uniaxial strain. (b) RISB quasiparticle dispersions for weak hole (\(n=1.97\)) and electron (\(n=2.03\)) doping. The shaded regions indicate the energy window around the Fermi level in which only one spin sector is present. (c,d) Corresponding spin-resolved Fermi surfaces for \(n=1.97\) and \(n=2.03\), respectively. Uniaxial strain removes one spin sector from the Fermi surface, resulting in complete spin polarization of the low-energy carriers. Reversing the sign of doping switches the spin polarization between the two spin sectors while the underlying altermagnetic order remains nearly compensated.} \label{fig4}
\end{figure}	
The pronounced sensitivity of the low-energy electronic structure to doping suggests that the effect of uniaxial strain should be particularly strong in the vicinity of half-filling, where relatively small changes of the dispersion can determine whether a given spin-split band crosses the Fermi level. We therefore focus in Fig.~\ref{fig4} on weak hole and electron doping, \(n=1.97\) and \(n=2.03\), respectively, in the presence of uniaxial strain. While the altermagnetic order and substantial quasiparticle renormalization remain present [Fig.~\ref{fig4}(a)], strain value of about 1$\%$ lifts the equivalence of the symmetry-related momentum directions and strongly reconstructs the low-energy bands. Most importantly, for both signs of doping the Fermi level intersects states belonging to only one of the two spin sectors over the entire Fermi surface [Fig.~\ref{fig4}(b)]. This is particularly transparent in the Fermi-surface maps of Figs.~\ref{fig4}(c,d): for hole doping, \(n=1.97\), only the cyan spin sector forms Fermi pockets, whereas for electron doping, \(n=2.03\), only the magenta sector remains at the Fermi level. The carrier spin polarization therefore reverses upon changing from hole to electron doping. Thus, the combined action of weak doping and uniaxial strain converts the correlated altermagnetic metal into a fully spin-polarized metallic state, while retaining the nearly compensated nature of the underlying magnetic order. This provides a route to selecting the spin character of the conducting carriers simply through the sign of doping. Therefore, changing the carrier type for example by gating for a given strain will reverse the spin polarization of the Fermi surface.

\subsection{Bilayer structure}

While corrrelation-assisted strain modifies effectively the in-plane hopping anisotropy in the monolayer, a bilayer offers an alternative route to controlling altermagnetism through the symmetry and strength of the interlayer hybridization.
%
\begin{figure}[t]
      \includegraphics[width=8.5cm]{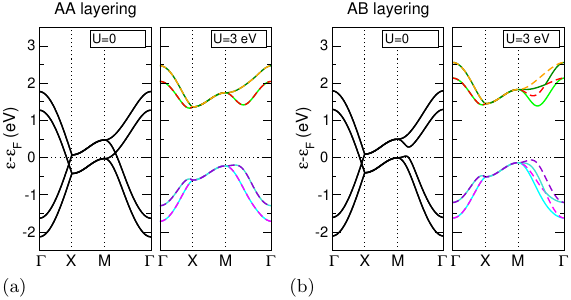}
      \caption{ Calculated stacking dependence of the bilayer altermagnetic electronic structure within RISB for $t_\perp=-0.25$\,eV . (a) Electronic dispersions along the high-symmetry path \(\Gamma-X-M-\Gamma\) for $AA$ stacking at \(U=0\) (PM) and \(U=3)\,\mathrm{eV}\) (AFM). (b) Corresponding dispersions for $AB$ stacking. In the $AB$ configuration, the dominant interlayer hopping connects opposite sublattices and favors parallel (ferroic) alignment of the layer Néel vectors, such that the momentum-dependent altermagnetic spin splitting of the individual layers combines constructively. In contrast, the $AA$ stacking suppresses the bilayer altermagnetic splitting. The comparison identifies $AB$ stacking as the configuration favorable for robust altermagnetism} \label{figb1}
\end{figure}	
We consider the two representative configurations shown in Fig.~\ref{figb1}: $AA$ and $AB$ stacking as shown in Fig.\ref{fig1}. In both configurations, $AA$ and $AB$, the interlayer hybridization generates sizeable interlayer AFM exchange coupling, i.e. bonding-antibonding splitting with stronger odd spin susceptibility, which favors antiferromagnetic order between the layers. Within a localized spin picture it refers to an AFM  $J_\perp\sim  t^2_\perp/U$. Then for the $AB$ stacking, \(t_\perp\) connects opposite sublattices of neighboring layers. The antiferromagnetic exchange associated with this hopping therefore favors antiparallel moments across an individual interlayer bond, which, because the intralayer order itself is staggered, corresponds to a parallel (ferroic) alignment of the Néel vectors of the two layers, \(\mathbf{L}_1=\mathbf{L}_2\). The momentum-dependent altermagnetic spin splitting generated by the two layers consequently adds constructively. In contrast, when the dominant interlayer hopping connects equivalent sublattices, the same antiferromagnetic interlayer coupling favors an antiparallel alignment of the layer Néel vectors, \(\mathbf{L}_1=-\mathbf{L}_2\), thereby opposing the cooperative buildup of the bilayer altermagnetic splitting. This stacking dependence is reflected in the electronic dispersions shown in Fig.~\ref{figb1}. Starting from the corresponding noninteracting bilayer bands at \(U=0\), the interacting RISB solution at \(U=3\,\mathrm{eV}\) develops a pronounced momentum-dependent spin splitting for the $AB$ configuration, while this is absent in the $AA$-stacking. Thus, the stacking geometry does not merely modify the bonding--antibonding band structure, but controls how the altermagnetic order parameters of the individual layers combine. In the following, we therefore focus on the $AB$ stacking as the configuration favorable for robust bilayer altermagnetism.
%
\begin{figure}[t]
      \includegraphics[width=8.5cm]{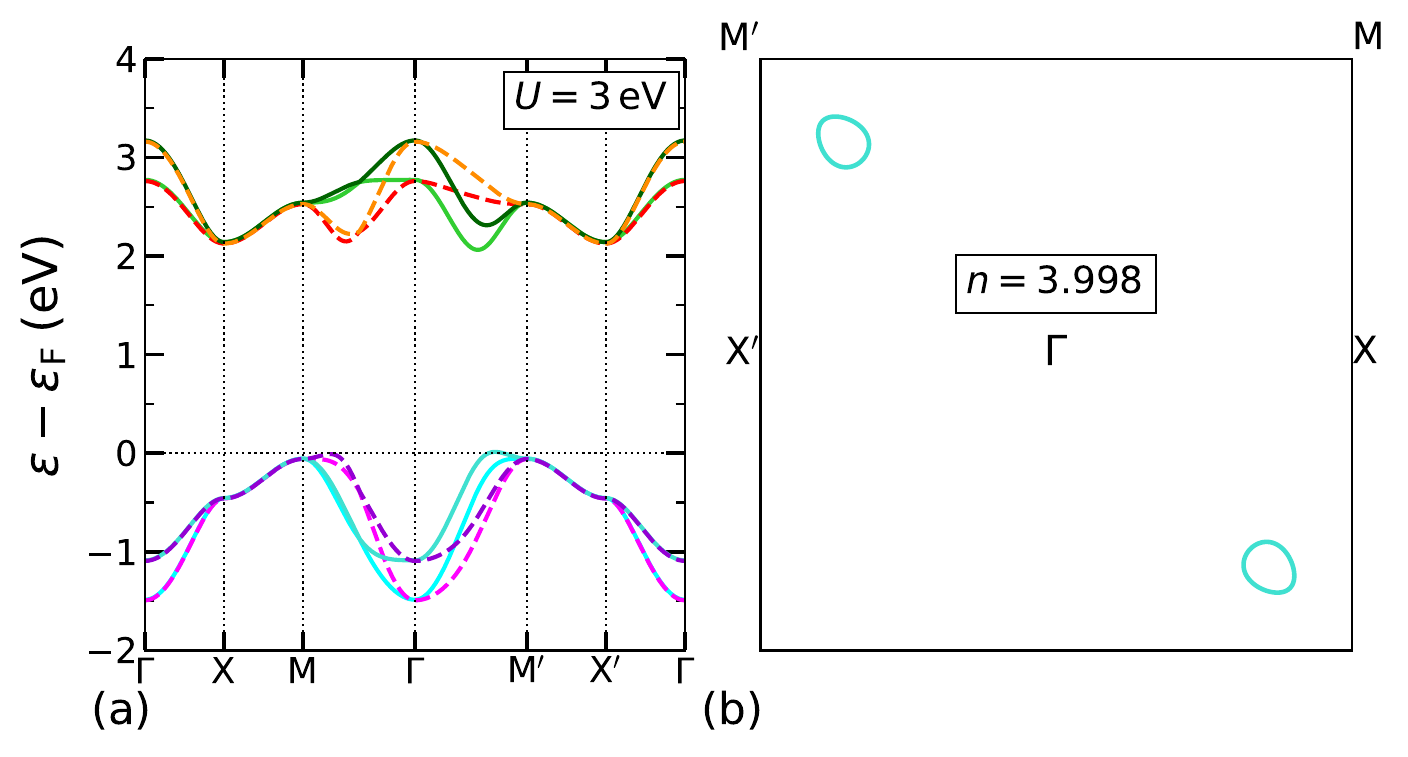}
      \caption{Calculated doping-induced fully spin-polarized Fermi surface in the $AB$-stacked bilayer as a result of doping and weak intra-unit cell checkerboard charge order. (a) Spin-resolved quasiparticle dispersion of the $AB$-stacked bilayer at \(U=3\,\mathrm{eV}\) and weak hole doping, \(3.998\). The ferroic alignment of the altermagnetic order parameters in the two layers and the spontaneosuly developed intraunitcell charge order produces a pronounced spin splitting of the low-energy band edges, such that the Fermi level intersects states of only one spin sector. (b) Corresponding spin-resolved Fermi surface, showing the resulting complete spin polarization of the low-energy carriers. The state remains nearly magnetically compensated with small net ferromagnetic moment.} \label{figbi2}
\end{figure}	

In particular, the constructive combination of the altermagnetic spin splitting in the $AB$-stacked configuration has an important consequence for the low-energy electronic structure. Calculating the altermagnetic order self-consistently we find that in the case of non-equivalent doping of two layers (modeled by the small difference in crystal-field splitting of A and B layers), an additional intraunit-cell charge density modulations develops, yielding a weak ferromagnetic component.
In particular, Figure~\ref{figbi2} demonstrates the resulting evolution of the $AB$-stacked bilayer upon inequavalent weak hole doping. Remarkably, already for a small deviation from half-filling, \(3.998\), the chemical potential intersects states belonging to only one spin sector, resulting in a fully spin-polarized Fermi surface with ferromagnetic component. In addition, the resulting charge density wave yield the breaking of the $C_4$ symmety, leaving only two Fermis urface pockets, crossing the Fermi level, see Fig.~\ref{figbi2}(b). This realizes a particularly interesting regime in which the low-energy charge carriers are again completely spin polarized. Remarkably, in the $AB$-stacked bilayer this regime is reached by weak layer-asymmetric doping and the associated charge-order instability, without requiring the external uniaxial strain used in the monolayer. Thus, layering provides an alternative route to controlling the spin polarization of an altermagnetic metal without requiring an external symmetry-breaking strain.

\begin{figure*}[htp!]
      \includegraphics[width=0.99\linewidth]{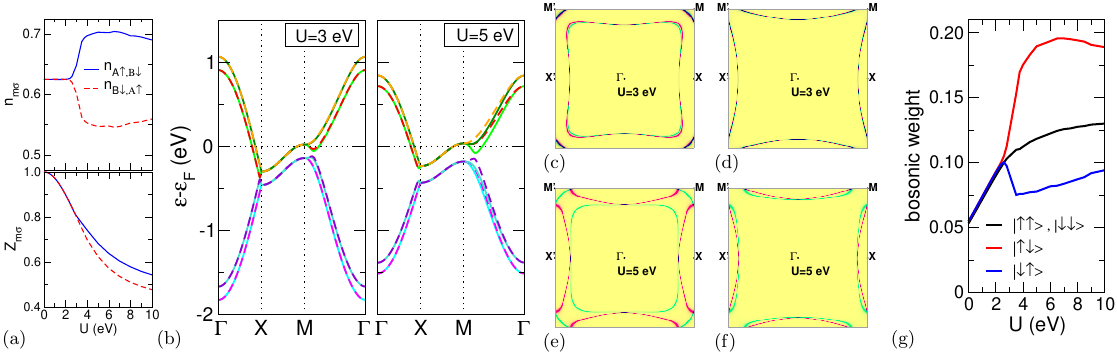}
      \caption{ Calculated correlation-driven evolution of the $AB$ bilayer altermagnetic state within RISB at interger filling $n=5$ for $t_\perp=-0.1$\,eV. (a) Spin-resolved quasiparticle weights \(Z_{m\sigma}\) and spin-sublattice occupations as a function of the Hubbard interaction \(U\). The altermagnetic order develops continuously above \(U_c\), whereas the quasiparticle weight undergoes a substantially stronger suppression at larger \(U\), indicating an evolution toward a strongly correlated, Mott-like regime. (b) RISB quasiparticle dispersions at \(U=3\,\mathrm{eV}\) and \(U=5\,\mathrm{eV}\), illustrating the interaction-induced altermagnetic spin splitting and increasing quasiparticle renormalization. (c,d) Spin-resolved Fermi surfaces at \(U=3\,\mathrm{eV}\). (e,f) Corresponding Fermi surfaces at \(U=5\,\mathrm{eV}\), demonstrating the correlation-induced reconstruction of the low-energy electronic structure. (g) RISB bosonic weights $\langle\phi(\sigma_A,\sigma_B)\rangle^2$ of the two-particle
      cluster states $|\sigma_A\sigma_B\rangle=|\uparrow\uparrow\rangle,|\downarrow\downarrow\rangle,|\uparrow\downarrow\rangle,|\downarrow\uparrow\rangle$ as a function of \(U\). The continuous splitting of the initially degenerate \(ud\) and \(du\) weights accompanies the onset of altermagnetic order, while the finite weight of several local configurations illustrates the many-body character of the correlated ordered state.} \label{figbi3}
\end{figure*}

Finally we would like to point out that the altermagnetic transition is not always Slater-like within RISB. For the integer filling of $n=5$ of the $A$B bilayer, a much richer behavior is revealed. In this scenario, five electrons have to be distributed among the two basic $AB$ clusters in the unit cell, and hence without breaking the layer symmetry this results in a metallic
solution even at strong coupling. Four of those electrons may be prone to singlet formation on the given clusters (as in the previously
discussed half-filled case), yet the remaining fifth electron will remain itinerant and scatter with these (nearly) localized singlets.
We observe altermagnetism with increasing interaction strength $U$ within such a challenging correlation scenario, as summarized in Fig.~\ref{figbi3}.

Observe here that the initially equivalent spin-sublattice occupations separate smoothly above a critical interaction \(U_c\), signaling the spontaneous development of compensated altermagnetic order out of a paramagnetic phase. At the same time, the transition occurs while the QP weight \(Z_{m\sigma}\) remains relatively large, demonstrating that the magnetic instability develops within a coherent metallic state rather. Upon further increasing \(U\), however, \(Z_{m\sigma}\) is substantially suppressed, displaying a much more pronounced correlation-induced renormalization than in the configurations discussed above. This progressive loss of QP weight indicates an evolution toward a strongly correlated, Mott-like regime, although \(Z_{m\sigma}\) remains finite over the interaction range considered here. The distinction between the onset of magnetic order and the subsequent strong QP renormalization is also reflected in the interacting dispersion. At \(U=3\,\mathrm{eV}\), shortly after entering the ordered regime, the characteristic momentum-dependent altermagnetic spin splitting is already clearly established, while at \(U=5\,\mathrm{eV}\) the stronger correlations substantially narrow and reconstruct the QP bands and the corresponding Fermi surfaces [Figs.~\ref{figbi3}(b)--(f)]. A complementary many-body perspective on this evolution is provided by using the bosonic degrees of freedom $\phi$ to inspect the two-particle cluster weights $\langle\phi(\sigma_A,\sigma_B)\rangle^2$, as shown in Fig.~\ref{figbi3}(g). In the paramagnetic state, the two oppositely oriented spin configurations $\uparrow\downarrow\rangle,|\downarrow\uparrow\rangle$ on the $AB$ clusters have equal weight. At \(U_c\), their degeneracy is lifted continuously, and their weights evolve in opposite directions upon entering the altermagnetic phase. The difference between these cluster weights therefore provides a local many-body measure of the developing staggered order and mirrors the continuous splitting of the spin-sublattice occupations. At the same time, the parallel-spin sector $\uparrow\uparrow\rangle,|\downarrow,\downarrow\rangle$ remains finite and evolves smoothly, demonstrating that the correlated altermagnetic state cannot be reduced to a single classical spin configuration but retains substantial contributions from several local many-body states. The simultaneous evolution of the multiplet weights and \(Z_{m\sigma}\) thus reveals two distinct aspects of the interaction-driven physics: a continuous redistribution of two-site configurations associated with spontaneous altermagnetic symmetry breaking at \(U_c\), followed at stronger coupling by a pronounced loss of quasiparticle weight and an approach toward a Mott-like correlated regime.

\section{Summary and Conclusions}

In summary, we have investigated how electronic correlations, doping, uniaxial strain, and interlayer coupling control the stability and low-energy electronic structure of altermagnetic states within a minimal Hubbard-type model. By comparing conventional Hartree--Fock theory with the rotationally invariant slave-boson approach, we have explicitly assessed the robustness of the mean-field description against correlation-induced quasiparticle renormalization. While RISB quantitatively modifies the critical interaction and substantially renormalizes the quasiparticle dispersion, the characteristic momentum-dependent spin splitting of the altermagnetic state remains robust. Altermagnetic order generally develops while the quasiparticle weight remains finite, demonstrating that the magnetic instability can occur within a correlated metallic regime rather than being tied to quasiparticle localization.

Away from half-filling, the RISB solution reveals a pronounced asymmetry between electron and hole doping, both in the quasiparticle renormalization and in the reconstruction of the spin-resolved Fermi surfaces. This sensitivity of the low-energy band structure to carrier concentration makes the altermagnetic metal particularly susceptible to symmetry-lowering perturbations. In the monolayer, we find that the combined action of weak doping and uniaxial strain can remove one spin sector entirely from the Fermi surface. Remarkably, the surviving spin sector is reversed upon changing from hole to electron doping. The resulting state therefore supports completely spin-polarized low-energy carriers while retaining the compensated character of the underlying magnetic order.

The bilayer provides a complementary route to controlling this behavior through interlayer hybridization. We find that the relative stacking of the layers is crucial, since it determines how the altermagnetic order parameters of the individual layers combine. 
In the favorable $AB$ configuration, the dominant interlayer hopping connects opposite sublattices. The corresponding interlayer coupling favors a ferroic alignment of the layer Neel vectors. Upon weak layer-asymmetric doping, the layers become slightly inequivalent and an additional intra-unit-cell charge-density-wave order develops, such that the magnetic moments in the upper and lower layers no longer compensate completely. As a result, a small shift of the chemical potential away from half-filling is sufficient to access strongly spin-selective band edges. In particular, weak hole doping produces a Fermi surface formed by a single spin sector with small net ferromagnetic component and breaking the $C_4$ symmetry. Thus, in contrast to the monolayer, where strain and doping act cooperatively, the bilayer geometry can provide the required electronic reconstruction through interlayer coupling, slightly nonequivalent doping and spontaneous generation of the inraunit cell charge density wave order. Experimentally, this result can be achieved by gating of semiconducting altermagnets. 

Finally, at the alternative integer filling \(n=5\) of the AB bilayer, the RISB treatment uncovers a richer interaction-driven evolution by revealing
an interaction-driven continuous paramagnet-to-altermagnet transition. Here the altermagnetic order develops continuously at a critical interaction \(U_c\), whereas the QP weight undergoes a much stronger suppression at larger \(U\). The two effects therefore occur on distinct interaction scales: spontaneous altermagnetic symmetry breaking takes place while coherent quasiparticles are still strong, followed by a progressive loss of QP weight and an evolution toward a strongly correlated, Mott-like regime. The RISB cluster weights provide a complementary microscopic view of this behavior. The two oppositely oriented cluster spin configurations, which carry equal weight in the paramagnetic state, become continuously nonequivalent upon entering the altermagnetic phase, directly reflecting the development of staggered magnetic order. At the same time, several local configurations retain finite weight throughout the ordered phase, highlighting the many-body character of the correlated altermagnetic state beyond a static Hartree--Fock description.

Taken together, our results identify quasiparticle renormalization, strain, doping, and interlayer coupling as complementary microscopic control parameters for altermagnetic metals. A central outcome is that complete spin polarization of the low-energy carriers does not require a sizable ferromagnetic moment: it can emerge from the momentum-dependent spin splitting of an altermagnetic state combined with a suitable reconstruction of its Fermi surface. In the monolayer this reconstruction can be induced by strain and doping, whereas in the bilayer the same functionality can arise from the constructive combination of the layer-resolved altermagnetic order through an appropriate stacking geometry. These results establish correlated mono- and bilayer altermagnets as a versatile setting for realizing and controlling highly spin-polarized metallic states without conventional ferromagnetism.

\section{Acknowledgements} 
Computations were performed at the Ruhr-University Bochum. 

\appendix

\bibliography{literatur}
\end{document}